# Transverse momentum and pseudorapidity dependence of particle production in Xe-Xe collision at $\sqrt{s_{NN}} = 5.44$ TeV

Zhang-Li Guo, Bao-Chun Li and Hong-Wei Dong

[1]College of Physics and Electronics Engineering, State Key Laboratory of Quantum Optics and Quantum Optics Devices, Shanxi University, Taiyuan 030006, China

[2]Collaborative Innovation Center of Extreme Optics, Shanxi University, Taiyuan 030006, China

**Abstract:** Through the collision-system configuration, the Tsallis statistics is combined with a multisource thermal model. The improved model is used to investigate the transverse momentum and pseudorapidity of particles produced in Xe-Xe collisions at $\sqrt{s_{NN}} = 5.44$ TeV. We discuss detailedly the thermodynamic properties, which are taken from the transverse momentum $p_T$ distributions of $\pi$, $K$ and $p$ for different centralities. The pseudorapidity $\eta$ spectra of charged particles for different centralities are described consistently in the model. And, the model result can estimate intuitively the longitudinal configuration of the collision system.

## 1. Introducion

The important goal of the ultrarelativistic heavy-ion collisions is to find and study the Quark–Gluon Plasma (QGP), which is a new matter state of strongly interacting quarks and gluons at high temperature and density. From 2010 to 2019, the Large Hadron Collider (LHC) has mainly carried out *p-p*, *p*-Pb and Pb-Pb ion collision experiments at various collision energies, which can provide different types of collision-system configurations. In 2017, the LHC performed a different kind of hadron collision at high energy, i.e. the first Xe[129] ion collisions at $\sqrt{s_{NN}} = 5.44$ TeV [1-4]. Since the nucleons of the Xe[129] nucleus is fewer than that of Pb[208] nucleus, the investigation of Xe-ion collisions can roughly bridge or connect the gap between *p* and Pb ion collisions. As a good intermediate-size system, the Xe–Xe colliding system brings a wonderful opportunity to discuss the colliding-system size dependence of multiparticle production in high-energy nuclear collisions [5, 6]. The nucleus collisions at high energies offer numerous experimental data about charged particle production, such as pions, kaons and protons. The particle production in the collision contains the interaction effects between hard and soft QCD processes. The feature discussion of the particle distribution can be used to infer the evolution and dynamics of different collision systems at different center of mass energies.

With respect to the final-state observables in these collisions, the particle transverse momentum and pseudorapidity multiplicity are two key measurements to understand the particle-production process and the matter evolution under the extreme conditions. The transverse momentum spectra are very important because they can provide essential information about QGP created in the collisions. The charged-particle pseudorapidity multiplicity is related to the early geometry of the collision system and is of great interest to investigate the properties of the collision-system evolution. Recently, the ALICE Collaboration measured charged-particle transverse momentum spectra and multiplicity density in Xe–Xe collisions at $\sqrt{s_{NN}}=5.44$ TeV at the LHC [2, 3]. In this work, the transverse momentum spectra are analyzed in an improved multisource thermal model, where the Tsallis statistics is imported [7-9]. Combined with the collision picture, we also discuss the charged–particle pseudorapidity density for different collision centralities. The investigation of the particle production in different collision systems can help us understand the matter evolution in the different collisions.

## 2. The particle spectra in the improved multisource thermal model

In high-energy nucleon or nuclei collisions, the thermodynamic information of the system evolution is very rich. These identified particles produced in the collisions may be regarded as a multiparticle system. The identified particles emit from different sources. We can assume that many emission sources are formed in the interacting system [10, 11]. In the stationary reference frame of a considered source, the distribution function of the particle momentum $p^{'}$ is given by

$$f_{p^{'}}(p^{'}) = \frac{1}{N}\frac{dN}{dp^{'}} = Cp^{'2}[1+(q-1)\frac{\sqrt{p^{'2}+m_0^2}-\mu}{T}]^{-q/(q-1)} \quad (1)$$

where $C$, $T$, and $q$ is the normalization constant, the temperature and the nonequilibrium degree parameter, respectively. The $q$ value is close to 1. For the chemical potential $\mu = 0$, the distribution function is

$$f_{p'}(p') = \frac{1}{N}\frac{dN}{dp'} = Cp'^{2}[1+(q-1)\frac{m_T}{T}]^{-q/(q-1)} \qquad (2)$$

When $q$ tends to 1, the density function is the standard Boltzmann distribution. The particle momentum function $p' = g(R_1)$ can be obtained by the Monte Carlo calculation, $\int_0^{p'} f(p')\,dp' < R_1 < \int_0^{p'+dp'} f(p')\,dp'$. The particle pseudorapidity $\eta'$ is

$$\eta' = \frac{1}{2}\ln\frac{E'+p_z'}{E'-p_z'}, \qquad (3)$$

where $E'$ and $p_z' = p'\cos\theta'$ is the energy and longitudinal momentum, respectively. The transverse momentum is

$$p_T' = \sqrt{p_x'^2 + p_y'^2} = p'\sin\theta', \qquad (4)$$

where $\theta' = \arctan[\frac{2\sqrt{r_2(1-r_2)}}{1-2r_2}]$ is the particle emission angle and is calculated by the Monte Carlo method. Due to $p_T = p_T'$, the distribution function of the particle transverse-momentum in the laboratory reference system frame is

$$f_{p_T}(p_T) = \frac{1}{N}\frac{d^2N}{dp_T d\eta} = C_T p_T m_T \cosh\eta'[1+(q-1)\frac{m_T \cosh\eta'}{T}]^{-q/(q-1)}. \qquad (5)$$

In contrast to the transverse momentum, the particle pseudorapidity $\eta$ in the laboratory reference system frame is not easy to calculate. Since $\eta'$ is a result of the source reference frame, one source is only considered in Eq. (1). For the calculation of the pseudorapidity, the space scale of the collision system cannot be ignored at the pseudorapidity $\eta$ space. Along the beam, these sources can be grouped into four categories as follow: a projectile leading-particle source with a pseudorapidity shift $\eta_{plp}$, a projectile cylinder composed of a series of sources with pseudorapidity shifts $\eta_{pc}$ ($\eta_{pc}^{\min} \leq \eta_{pc} \leq \eta_{pc}^{\max}$), a target cylinder composed of a series of sources with pseudorapidity shifts $\eta_{tc}^{min} \leq \eta_{tc} \leq \eta_{tc}^{max}$ and a target leading-particle source with a pseudorapidity shift $\eta_{tlp}$. In the laboratory reference system frame, the Monte Carlo pseudorapidity function of particles from the four parts can be written as

$$\eta_1 = \eta_{plp} + \eta', \tag{6}$$

$$\eta_2 = \eta_{pc} + \eta', \quad \eta_{pc}^{min} \leq \eta_{pc} \leq \eta_{pc}^{max}, \tag{7}$$

$$\eta_3 = \eta_{tc} + \eta', \quad \eta_{tc}^{min} \leq \eta_{tc} \leq \eta_{tc}^{max}, \tag{8}$$

$$\eta_4 = \eta_{tlp} + \eta'. \tag{9}$$

By the $p'$ distribution function Eq. (1), we can obtain the source pseudorapidity $\eta'$ in the stationary reference frame. Then, the pseudorapidity distribution in the laboratory reference frame can be derived from the $\eta$ space scale of the collision system, which is described by the collision Eq. (8)-(9).

## 3. Discussions and conclusions

Figure 1 shows transverse momentum $p_T$ distributions of pions π, kaons K and protons $p$ produced in Xe-Xe collisions at $\sqrt{s_{NN}} = 5.44$ TeV. The filled circles indicate the experimental data [2] for nine centrality bins (from 0–5% central collisions to 70–80% peripheral collisions). The lines are the results of the Eq. (5). For pions, kaons and protons, the nonequilibrium degrees are $q = 1.141$, $q = 1.080$ and $q = 1.022$, respectively. For the same particles, the $q$ is a constant value in each interval of the centrality. This reflects nonequivalent excitation of the thermal sources of the three particles in the centrality classes. The temperatures for the three kinds of particles are shown in Table 1-3 with $\chi^2/\text{ndf}$ and increase with the increase of the collision centrality. The $p_T$ differential cross sections for different collision centralities are governed by the temperature $T$, where the reaction system freezes out and the considered particles will no longer interact. The particles at low $p_T$ region are more close to a thermal equilibrium and the particles at high $p_T$ region are more close to be produced in a hard scatterings, which is determined by pQCD [12, 13] .

From pions to protons, these particle masses affect the slope of the transverse momentum $p_T$ spectra. So, the temperature $T$ and nonequilibrium degree $q$ depend on the final-state particle mass. With increasing particle mass, the temperature $T$ increases generally and the nonequilibrium degree $q$ deceases. The mass dependence may originate from the deformed nuclei, Xe. With the matter produced in the collision moving at a finite velocity, the Lorentz-boost magnitude of the momentum distribution occurs obviously and is proportional to the particle mass. Therefore, the $q$ values of $\pi$, $K$ and $p$ systems are different. This shows how close the three systems are to the kinetic equilibrium.

Figure 2 shows pseudorapidity $\eta$ spectra of charged particles produced in Xe–Xe collisions at $\sqrt{s_{NN}} = 5.44$ TeV. The filled circles indicate the experimental data [3] for twelve centrality bins (from 0–2.5% central collisions to 80–90% peripheral collisions). The lines are the results of the Eq. (6)-(9). The heights of the pseudorapidity spectra exhibit strong centrality dependences. It is because the number of observed particles is approximately proportional to the number of collision participant nucleons, which is a function of the impact parameter. The configuration parameters of the thermalized cylinder are shown in Table 4. The $\eta_{pc}^{max}$ and $\eta_{pc}^{min}$ slightly increase with collision centralities. The pseudorapidity distributions of the peripheral collision are wider than that of the most central collision. So, the length of the thermalized cylinder at $\eta$ space decreases with the impact parameter. It means the number of thermal sources produced in Xe-Xe collision increases with centralities. The source contributions from different categories are seen intuitively and the configuration of the collision system is quantized visually. It helps us understand the influence of the collision-system size and the evolution information of the produced matter in the collision [14, 15].

In this paper, the Tsallis statistics is combined with the collision-system configuration, i. e. the multisource thermal model. We use the improved model to investigate the particle production in the intermediate-size collision system [16, 17], Xe-Xe collision. By the study of the transverse momentum $p_T$ distributions of $\pi$, $K$ and $p$, the temperature and nonequilibrium degree are obtained. The centrality dependence and the particle mass dependence are discusses. Based on the result, the pseudorapidity $\eta$ spectra of charged particles are reproduced. The configuration of the intermediate-size collision system is quantized visually by the collision picture, which can characterise the primary properties of the collision system.

## Acknowledgments

This work is supported by National Natural Science Foundation of China under Grants No. 11247250 and No. 11575103, Shanxi Provincial Natural Science Foundation under Grant No. 201701D121005, and Scientific and Technological Innovation Programs of Higher Education Institutions in Shanxi (STIP) Grant No. 201802017.

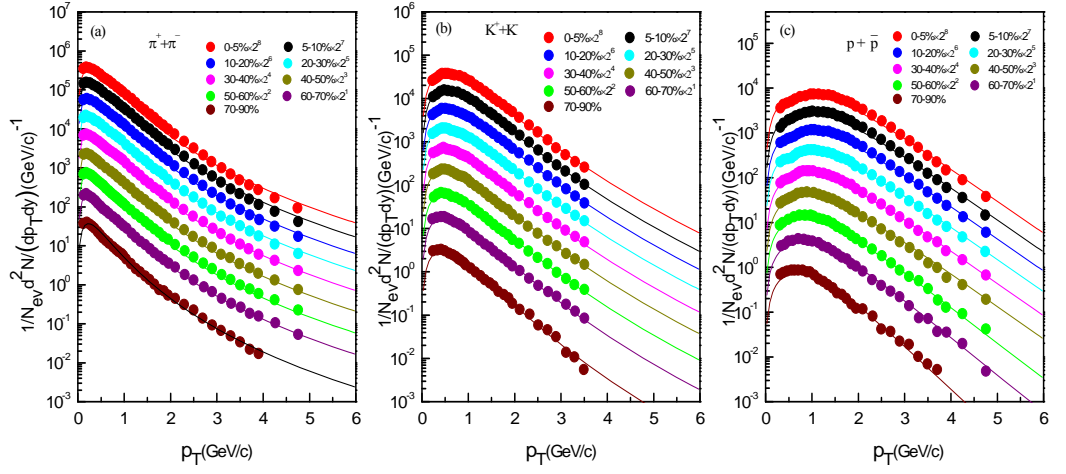

Fig 1. Transverse momentum distributions of π, K and p for nine centrality bins in Xe-Xe collisions at $\sqrt{s_{NN}}$ =5.44 Tev. The filled circles indicate the experimental data [2] for nine centrality bins. The lines are the results of the Eq. (5).

Table 1. Values of parameters used in Figure 1(a). The unit of $T$ is GeV.

| Centrality | $T$ | $\chi^2/\text{ndf}$ |
|---|---|---|
| 0-5% | 0.101 | 0.447 |
| 5-10% | 0.100 | 0.283 |
| 10-20% | 0.099 | 0.124 |
| 20-30% | 0.098 | 0.150 |
| 30-40% | 0.097 | 0.202 |
| 40-50% | 0.096 | 0.261 |
| 50-60% | 0.095 | 0.312 |
| 60-70% | 0.094 | 0.286 |
| 70-90% | 0.091 | 0.472 |

Table 2. Values of parameters used in Figure 1(b). The unit of $T$ is GeV

| Centrality | $T$ | $\chi^2/\text{ndf}$ |
|:---:|:---:|:---:|
| 0-5% | 0.202 | 0.165 |
| 5-10% | 0.200 | 0.160 |
| 10-20% | 0.199 | 0.144 |
| 20-30% | 0.198 | 0.105 |
| 30-40% | 0.196 | 0.275 |
| 40-50% | 0.195 | 0.369 |
| 50-60% | 0.191 | 0.424 |
| 60-70% | 0.183 | 0.571 |
| 70-90% | 0.166 | 0.601 |

Table 3. Values of parameters used in Figure 1(c). The unit of $T$ is GeV.

| Centrality | $T$ | $\chi^2/\text{ndf}$ |
|:---:|:---:|:---:|
| 0-5% | 0.382 | 0.317 |
| 5-10% | 0.381 | 0.295 |
| 10-20% | 0.379 | 0.210 |
| 20-30% | 0.378 | 0.226 |
| 30-40% | 0.377 | 0.305 |
| 40-50% | 0.374 | 0.514 |
| 50-60% | 0.342 | 0.590 |
| 60-70% | 0.324 | 0.646 |
| 70-90% | 0.278 | 0.675 |

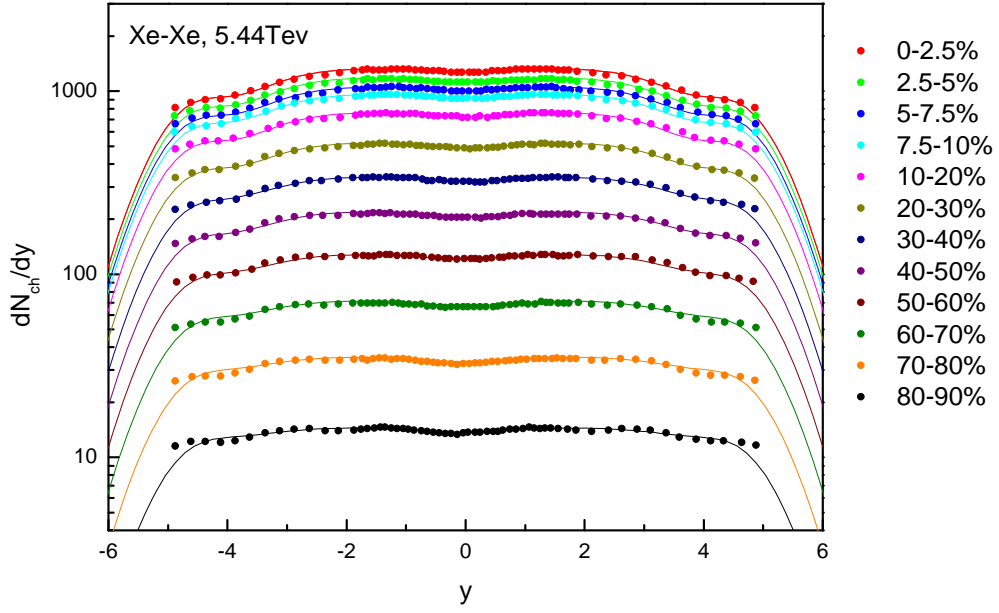

Fig 2. Pseudorapidity density of charged particles for 12 centrality bins in Xe–Xe collisions at $\sqrt{s_{NN}}$ =5.44Tev. The filled circles indicate the experimental data [3] for twelve centrality bins. The lines are the results of the Eq. (6)-(9).

Table 4. Values of parameters corresponding to the curves in Figure 2.

| Centrality | $\eta_{pc}^{max}$ | $\eta_{pc}^{min}$ | $\eta_{plp}$ | $k$ |
|---|---|---|---|---|
| 0-2.5% | 3.70 | 0.05 | 4.60 | 0.101 |
| 2.5-5% | 3.70 | 0.05 | 4.60 | 0.101 |
| 5-7.5% | 3.70 | 0.07 | 4.60 | 0.101 |
| 10-20% | 3.70 | 0.06 | 4.60 | 0.101 |
| 20-30% | 3.75 | 0.06 | 4.60 | 0.101 |
| 30-40% | 3.80 | 0.06 | 4.60 | 0.101 |
| 40-50% | 3.80 | 0.06 | 4.60 | 0.101 |
| 50-60% | 3.85 | 0.06 | 4.60 | 0.101 |
| 60-70% | 3.90 | 0.06 | 4.60 | 0.101 |
| 70-80% | 3.95 | 0.06 | 4.60 | 0.101 |
| 80-90% | 4.00 | 0.06 | 4.60 | 0.101 |